\documentclass[a4paper, 11pt]{article} 
\pdfoutput=1
\usepackage{amsmath}
\usepackage{amsthm}
\usepackage{amssymb}
\usepackage{amsfonts}
\usepackage{graphics}
\usepackage{epsfig}
\usepackage[colorlinks=true,bookmarks=false]{hyperref}
\graphicspath{{bilder/}}
\newtheorem{Def}{Definition}[section]
\newtheorem{Thm}[Def]{Theorem}
\newtheorem{Pro}[Def]{Proposition}
\newtheorem{Cor}[Def]{Corollary}

\newtheorem{Lem}[Def]{Lemma}
\newtheorem{Rem}[Def]{Remark}
\def\conv{\mathop{\mathrm{ conv}}\nolimits}
\def\inf{\mathop{\mathrm{ inf}}\nolimits}

\def\cos{\mathop{\mathrm{ cos}}\nolimits}     
\def\sin{\mathop{\mathrm{ sin}}\nolimits}     
\def\aff{\mathop{\mathrm{ aff}}\nolimits}     

\def\pos{\mathop{\mathrm{ pos}}\nolimits}
\def\rec{\mathop{\mathrm{ rec}}\nolimits}
\def\sup{\mathop{\mathrm{ sup}}\nolimits}

\def\inv{\mathop{\mathrm{ inv}}\nolimits}

\begin{document}
\vspace{-3cm}
\begin{center} 
\textbf{\Large{Polyhedral Voronoi Cells}} \\
\vspace{.3cm}
Ina Voigt\footnote{Fakult\"at f\"ur Mathematik, Technische 
Universit\"at Dortmund, Vogelpothsweg 87, D-44227 Dortmund, Germany, 
\texttt{ina.voigt@tu-dortmund.de}} \\
Stephan Weis\footnote{Department Mathematik, 
Friedrich-Alexander-Universit\"at Erlangen-N\"urnberg,
Bismarckstra{\ss}e 1$\frac{\text{1}}{\text{2}}$, D-91054 Erlangen, 
Germany, \texttt{weis@mi.uni-erlangen.de}} \\
\vspace{.1cm}
February 19, 2010
\end{center}
\noindent
{\small\textbf{\emph{Abstract --}}
Voronoi cells of a discrete set in Euclidean space are
known as generalized polyhedra. We identify polyhedral cells of a 
discrete set through a direction cone. For an arbitrary set we 
distinguish polyhedral from non-polyhedral cells using inversion at 
a sphere and a theorem of semi-infinite linear programming.
\\[1mm]
}
%
%
\section{Introduction}
\par
The Voronoi diagram of a finite set in the $n$-dimensional Euclidean 
space $\mathbb{E}^n$ is a popular concept in Discrete and 
Computational Geometry, cf.\ Aichholzer and Aurenhammer 
\cite{Aur02} or Okabe et al.\ \cite{OBS00}, as well as in Minkowski 
Geometry, cf.\ Section 4 in Martini and Swanepoel \cite{Mar04}.
\par
A natural generalization from a finite set is the concept 
of a discrete set. By definition, a subset of $\mathbb{E}^n$ is 
\emph{discrete} if its intersection with any bounded set of 
$\mathbb{E}^n$ is finite. Here $M\subset\mathbb{E}^n$ is 
\emph{bounded} 
if $\sup_{x,y\in M}\|x-y\|<\infty$ with the Euclidean norm 
$\|\,\cdot\,\|$ based on the Euclidean scalar product 
$\langle\,\cdot\,,\,\cdot\,\rangle$. Equivalently, a subset of 
$\mathbb{E}^n$ is discrete if it has no accumulation point.
\par
We study the cardinality of half spaces needed to describe a Voronoi 
cell. A \emph{closed half space} is defined for non-zero 
$u\in\mathbb{E}^n$ and $\lambda\in\mathbb{R}$  by
\[
H^-(u,\lambda)\;:=\;\{\,x\in\mathbb{E}^n\,\mid\,\langle x, 
u\rangle\leq\lambda\,\}.
\]
The \emph{Voronoi diagram} of a non-empty \emph{generator} 
$\mathcal{P}\subset\mathbb{E}^n$ is the tessellation 
of $\mathbb{E}^n$ consisting of the \emph{Voronoi cells}
\[ 
V(p)\; :=\; \{\, x \in \mathbb{E}^n
\,\mid\, \| x - p \| \leq \| x - q \| \; 
\text{ for all } q \in \mathcal{P}\, \},
\qquad p\in\mathcal{P}.
\]
By translational invariance we assume in this article that the origin 
$0_n$ of $\mathbb{E}^n$ belongs to $\mathcal{P}$ and we restrict to 
the cell $\mathcal{V}:=V(0_n)$ at $0_n$. Notice the closed half space 
representation
\begin{equation}
\label{intro:vorocellzero}
2\mathcal{V}\;=\;\textstyle\bigcap_{p \in \mathcal{P} \smallsetminus 
\{ 0_n \}}\,H^-(p,\| p \|^2),
\end{equation}
the intersection over the empty index set being understood as 
$\mathbb{E}^n$. 
\par
To catch the structure of a Voronoi cell of a discrete set we use the 
following definitions. Let a subset $M \subset\mathbb{E}^n$ be given.
The \emph{affine hull} $\aff(M)$ resp.\ \emph{positive hull} $\pos(M)$ 
of $M$ consists of sums $\lambda_1x_1+\cdots+\lambda_Nx_N$ such that
for $i=1,\ldots,N$ we have $\lambda_i\in\mathbb{R}$, $x_i\in M$ and 
$\lambda_1+\cdots+\lambda_N=1$ resp.\ $\lambda_i\geq 0$. Notice
$\aff(\emptyset)=\emptyset$ and $\pos(\emptyset)=\{0_n\}$. The 
\emph{convex hull} of $M$ is $\conv(M):=\aff(M)\cap\pos(M)$. If 
$M=\conv(M)$ then $M$ is \emph{convex}. If for any $\lambda\geq 0$ and 
$\lambda M:=\{\lambda m\mid m\in M\}$  we have $\lambda M\subset M$, 
then $M$ is a \emph{cone}. A convex cone is \emph{finitely generated} 
if it is the positive hull of a finite set. A \emph{polyhedron} is the 
intersection of finitely many closed half spaces. A bounded polyhedron 
is a \emph{polytope}. If $M$ is convex and if any intersection of $M$ 
with a polytope is a polytope, then $M$ is a
\emph{generalized polyhedron}.
\begin{Rem}\label{generalized_poly}\upshape
It is clear from (\ref{intro:vorocellzero}) that a Voronoi cell of a 
finite generator is a polyhedron. It is proved by Gruber 
\cite{Grub07}, Chapter 32, that a Voronoi cell of a discrete generator 
is a generalized polyhedron.
\end{Rem}
\par
\begin{figure}
\begin{center}
\begin{picture}(5.6,3.5)
\put(0,0){\includegraphics[height=3.5cm, bb=0 0 500 300]%
{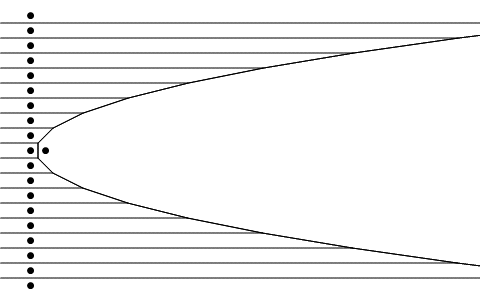}}
\put(4,2){$\mathcal{V}$}
\end{picture}
\caption{\label{fig:bsp3_2_1}The Voronoi diagram of the generator
$\mathcal{P} := \{ (-1,z) \mid z \in\mathbb{Z} \}\cup \{ 0_2 \}$ is 
depicted. The direction cone 
$\mathcal{D}=\{(x,y)\in\mathbb{R}^2\mid x<0\}\cup\{0_2\}$ is not 
closed, hence it is not finitely generated. Theorem~\ref{inas_theorem} 
concludes that the Voronoi cell $\mathcal{V}$ is not a polyhedron.}
\end{center}
\end{figure}

The existence of a non-polyhedral Voronoi cell for a 
discrete generator is demonstrated in Figure~\ref{fig:bsp3_2_1}. One 
of us has characterized a polyhedral Voronoi cell through the 
\emph{direction cone}\footnote{In optimization the cone $\mathcal{D}$ 
is called the \emph{cone of feasible directions}.} 
\[
\mathcal{D}\;:=\;\pos(\mathcal{P}).
\] 
\begin{Thm}[Ina Voigt~\cite{Voigt08}]
\label{inas_theorem}
If $\mathcal{P}$ is discrete then the Voronoi cell $\mathcal{V}$ 
is a polyhedron if and only if the direction cone $\mathcal{D}$ is 
finitely generated.
\end{Thm}
\par
Further examples to apply this theorem are in Figure~\ref{fig:hyp}
and Figure~\ref{fig:par}. For subsequent discussions we recall that a 
convex cone is finitely generated if and only if it is a polyhedron, 
cf.\ Ziegler \cite{Zie95}. If $F$ is a convex subset of a convex 
subset $C\subset\mathbb{E}^n$ such that for $x,y\in C$ and 
$0<\lambda<1$ the inclusion $\lambda x+(1-\lambda)y\in F$ always 
implies $x,y\in F$, then $F$ is a \emph{face} of $C$. The zero 
dimensional faces are the \emph{extreme points} and an
\emph{extreme ray} is a face which is a half line emanating from the 
origin. We notice that a  polyhedron has at most finitely many faces, 
cf.\ \S 19 in Rockafellar \cite{Rock72}.
\par
In this article we recover Theorem~\ref{inas_theorem} from results
about a generator $\mathcal{P}$ not necessarily discrete. As a tool
we use the diffeomorphism of inversion at the unit sphere
\[
\inv\;:
\quad\mathbb{E}^n\setminus\{0_n\}\,\longrightarrow\,
\mathbb{E}^n\setminus\{0_n\},\quad
x\,\longmapsto\,\frac{x}{\|x\|^2},
\]
the \emph{reciprocal} 
$\mathcal{R}:=\inv(\mathcal{P}\setminus\{0_n\})$ and the
\emph{convex reciprocal}
\[
\mathcal{C}\;:=\;\conv(\,\mathcal{R}\cup\{0_n\}\,).
\]
Illustrations of $\mathcal{C}$ are given in the following figures.
We apply in Section~\ref{sec:char_cone} a theorem of semi-infinite 
linear programming by Goberna and L\'opez and prove that the Voronoi 
cell $\mathcal{V}$ is a polyhedron if and only if the closure 
$\overline{\mathcal{C}}$ is a polyhedron. The intuition is that
$\mathcal{V}$ is completely surrounded by the Voronoi cells 
corresponding to the extreme points of $\overline{\mathcal{C}}$ while 
$\mathcal{V}$ has no extension in unbounded directions of 
$\overline{\mathcal{C}}$. The surrounding cells are finite in number 
only if $\overline{\mathcal{C}}$ is a polyhedron. In the case of a 
discrete generator $\mathcal{P}$ the condition relaxes to the 
condition that $\mathcal{C}$ is a polytope.
\par
Compared to the concept of convex reciprocal, the direction cone
\[
\mathcal{D}\;=\;\pos(\mathcal{P})
\;=\;\pos(\mathcal{R})
\]
is nearer to the geometry of the generator $\mathcal{P}$. We think 
of $\mathcal{D}$ as the area $0_n+\mathcal{D}$ occupied by 
$\mathcal{P}$ from point of view of $0_n$. For arbitrary $\mathcal{P}$
we find in Section~\ref{sec:direction_cone} that a polyhedral closure 
$\overline{\mathcal{D}}$ is necessary for a polyhedral Voronoi cell 
$\mathcal{V}$, i.e.\ a finite number of extreme rays of 
$\overline{\mathcal{D}}$ is necessary. On the other hand, a polyhedral 
$\mathcal{D}$ is not sufficient, a simple example being 
Figure~\ref{fig:kreis}.
\par
Stronger conditions apply to the case of a discrete generator 
$\mathcal{P}$, we recover Theorem~\ref{inas_theorem}. The generator 
must be enclosed in a half space $H^-(u,0)$ for non-zero 
$u\in\mathbb{E}^n$ to realize a non-polyheral Voronoi cell 
$\mathcal{V}$. Then any asymptotic direction of the generator not 
in $\mathcal{D}$, i.e.\ any accumulation point of
$\{\frac{x}{\|x\|}\mid x\in\mathcal{P}\setminus\{0_n\}\}$ not 
in $\mathcal{D}$, makes a finite half space representation of 
$\mathcal{V}$ impossible, see Figure~\ref{fig:bsp3_2_1} 
and~\ref{fig:hyp} for examples. (For the general generator there is no 
such condition, see Figure~\ref{fig:klappe} as a counterexample.) 
\par
In Section~\ref{sec:bounded_cells} we discuss the polar $\mathcal{D}^*$
of the direction cone $\mathcal{D}$. This is the normal cone at $0_n$ 
of the convex hull $\conv(\mathcal{P})$ of the generator,
\[
\mathcal{D}^*\;=\;
\{\,x\in\mathbb{E}^n\, \mid\, \langle x,y\rangle\leq 0
\text{ for all }y\in\conv(\mathcal{P})\,\}.
\]
As we noted above, the cone $\overline{\mathcal{D}}=\mathcal{D}^{**}$ 
is useless as a sufficient condition for a polyhedral Voronoi cell. 
However, it is useful to decide if a Voronoi cell is bounded. This 
problem is resolved in the literature for finite generators, see e.g.\ 
Okabe et al.\ \cite{OBS00}.
%
%
%
%
%
%
%
%
\section{The characteristic cone}
\label{sec:char_cone}
\par
We apply a theorem from semi-infinite linear programming to the
special case of the Voronoi cell $\mathcal{V}$ and obtain conditions 
on the cone of inequalities for the half spaces representation of 
$\mathcal{V}$. The result is interpreted in terms of the convex 
reciprocal $\mathcal{C}$. The cone of inequalities is well-known from 
the lifting construction for Delaunay triangulations. 
\par
The starting point is the, possibly infinite, system of linear 
inequalities 
\[
\textstyle
\sigma\;:=\;\left\{\,\langle p,x \rangle \leq \| p \|^2
\,\mid\,p\in\mathcal{P}\,\right\}
\]
satisfied by an unknown $x\in\mathbb{E}^n$, if and only if $x$ belongs 
to the doubly sized Voronoi cell $2\mathcal{V}$, see 
(\ref{intro:vorocellzero}). The \emph{characteristic cone} of 
$\sigma$ is\footnote{In \cite{Gob98} the characteristic cone is used 
with the opposite sign (reflected at the origin) compared to our 
definition.}
\[
\textstyle
K\;:=\;\pos\left(\,\left\{\,
\left(\begin{smallmatrix}p\\\|p\|^2\end{smallmatrix}\right)
\,\mid\,
p\in\mathcal{P}\setminus\{0_n\}\,\right\}\,\cup\,\left\{\,
\left(\begin{smallmatrix}0_n\\1\end{smallmatrix}\right)
\,\right\}\,\right)\;\subset\;\mathbb{E}^{n+1}.
\]
The trivial equation for $0_n\in\mathcal{P}$ is omitted. As a special 
case of Theorem 5.13 in Goberna and L\'opez \cite{Gob98} the 
following equivalence holds.
\begin{Thm}[Goberna and L\'opez]
\label{gobyl}
$\mathcal{V}$ is a polyhedron if and only if the 
closure $\overline{K}$ of the 
characteristic cone $K$ is a polyhedron.
\end{Thm}
\par
In place of $K\subset\mathbb{E}^{n+1}$ we can study the convex 
reciprocal 
\[
\textstyle
\mathcal{C}\;=\;\conv(\,\{\,x\,\|x\|^{-2}\,\mid\,x\in\mathcal{P}
\setminus\{0_n\}\,\}\;\cup\;\{0_n\}\,)
\;\subset\;\mathbb{E}^n.
\]
We have
$K=\bigcup_{\lambda\geq 0}\,\lambda\,(\mathcal{C}\times\{1\})$ and 
$\mathcal{C}\times\{1\}=K\cap\,\{x_{n+1}=1\}$. We must be careful in a 
discussion of closures: if $\mathcal{C}$ is unbounded then 
$\overline{K}$ has non-zero points in the hyperplane $x_{n+1}=0$ but
$K$ does not.
\begin{Pro}
\label{pro:gobyl}
The following statements are equivalent:
\begin{enumerate}
\item[(i) ] $\mathcal{V}$ is a polyhedron,
\item[(ii) ] $\overline{K}$ is a polyhedron,
\item[(iii) ] $\overline{\mathcal{C}}$ is a polyhedron.
\end{enumerate}
\end{Pro}
\par\noindent{\bf Proof.} 
We recall from Theorem 11.5 in Rockafellar \cite{Rock72} that a closed 
convex set is the intersection of the closed half spaces that contain 
the set. More specific assertions about closures of convex hulls 
and closures of positive hulls are Corollary 11.5.1 and Corollary 
11.7.2 in the same reference. As a consequence we can write for the 
same (possibly empty) index set 
$I\subset\mathbb{E}^n\setminus\{0_n\}\times\mathbb{R}$
\[
\textstyle
\overline{\mathcal{C}}\;=\;
\bigcap_{(u,\lambda)\in I}H^-(u,\lambda)
\]
and
\[
\textstyle
\overline{K}\;=\;H^-\left(
\left(\begin{smallmatrix}0_n\\-1\end{smallmatrix}\right),0\right)
\;\cap\;\bigcap_{(u,\lambda)\in I}
H^-\left(\left(\begin{smallmatrix}u\\-\lambda\end{smallmatrix}\right),
0\right).
\]
Notice that the closed half spaces for the convex cone $\overline{K}$ 
have the origin $0_{n+1}$ on their boundary hyperplane. A particular 
result is that $\overline{K}$ and $\overline{\mathcal{C}}$ are both 
polyhedra or they are both not. Theorem~\ref{gobyl} completes the 
proof.
\hfill $\Box$ \\
\par
\begin{figure}
\begin{center}
\begin{picture}(12,3.5)
\put(0,0){\includegraphics[height=3.5cm, bb=0 0 500 300, clip=]%
{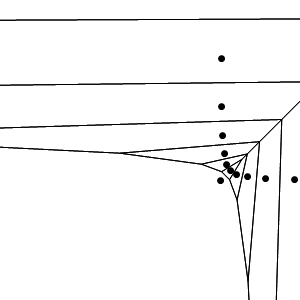}}
\put(8.4,0){\includegraphics[height=3.5cm, bb=0 0 500 300, clip=]%
{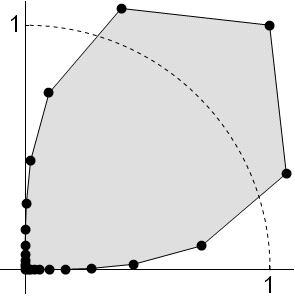}}
\put(1.0,0.5){$\mathcal{V}$}
\put(9.8,1.4){$\mathcal{C}$}
\end{picture}
\caption{\label{fig:hyp}
$\mathcal{P}:=\{\frac{1}{2}(\cosh(x)+\sinh(x),\cosh(x)-\sinh(x))%
\mid x\in\frac{1}{2}\mathbb{Z}\}\cup\{0_2\}$ is depicted with its
Voronoi diagram (left). The Voronoi cell $\mathcal{V}$ is not a 
polyhedron by Theorem~\ref{inas_theorem} as 
$\mathcal{D}=\{(x,y)\in\mathbb{E}^2\mid x,y>0\}\cup\{0_2\}$ is not 
finitely generated. A second argument that $\mathcal{V}$ is not a 
polyhedron is Corollary~\ref{cor2:gobyl} because the convex reciprocal 
$\mathcal{C}$, having the infinitely many extreme points 
$\mathcal{R}$, is not a polytope (right).}
\end{center}
\end{figure}
\begin{figure}
\begin{center}
\begin{picture}(12,3.5)
\put(0,0){\includegraphics[height=3.5cm, bb=0 0 500 300, clip=]%
{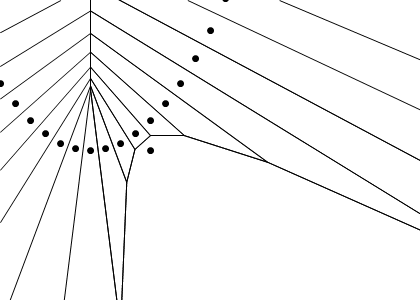}}
\put(6.5,0){\includegraphics[height=3.5cm, bb=0 0 600 300, clip=]%
{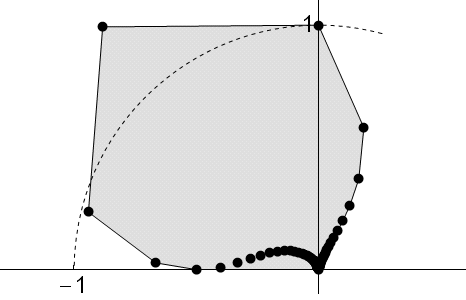}}
\put(3,0.5){$\mathcal{V}$}
\put(9.1,1.5){$\mathcal{C}$}
\end{picture}
\caption{\label{fig:par}The Voronoi diagram of
$\mathcal{P}:=\{(t,f(t))\mid t\in\frac{1}{2}\mathbb{Z}\}%
\cup\{0_2\}$ is depicted (left). 
The graph of the real function $f:t\mapsto\frac{1}{4}(t+2)^2$ has 
tangents through $(-2,0)$ and $(2,4)$ meeting $0_2$. Thus, there is a 
steepest line through $0_2$ that meets a point of
$\mathcal{P}\setminus\{0_2\}$ so the direction cone $\mathcal{D}$ is 
finitely generated. Then Theorem~\ref{inas_theorem} proves that 
$\mathcal{V}$ is a polyhedron. A second argument that $\mathcal{V}$ is 
a polyhedron is Corollary~\ref{cor2:gobyl}. The two branches
$\inv(t, f(t))$ are concave near $0_2$ (for large $|t|$), so almost 
all points of $\mathcal{R}$ belong to the interior of $\mathcal{C}$, 
which therefore is a polytope (right).}
\end{center}
\end{figure}

Proposition~\ref{pro:gobyl} is explained 
with non-discrete examples in Figure~\ref{fig:kreis} and
Figure~\ref{fig:klappe}. The statement of the proposition simplifies 
in the discrete case with the following remark. This is stated in 
Corollary~\ref{cor2:gobyl} and explained with two discrete examples in 
Figure~\ref{fig:hyp} and Figure~\ref{fig:par}.
\begin{Rem}\upshape
\label{rem:compact}
If $\mathcal{P}$ is discrete then the convex reciprocal 
$\mathcal{C}$ is compact and the characteristic cone $K$ is 
closed. Observe that the reciprocal $\mathcal{R}$ is bounded having
$0_n$ as the only possible accumulation point. Then
$\mathcal{R}\cup\{0_n\}$ is compact and from Carath\'eodory's theorem 
follows that 
$\mathcal{C}=\conv(\mathcal{R}\cup\{0_n\})$ is compact. Under the 
linear map $\alpha:\;\mathbb{E}^{n+1}\to\mathbb{E}^{n+1},\;(x,\lambda)
\mapsto(\lambda x,\lambda)$, the closed cylinder
$\widetilde{\mathcal{C}}:=\mathcal{C}
\times\{\lambda\geq 0\}\subset\mathbb{E}^{n+1}$ 
is mapped to $K$.  The kernel $\mathbb{E}^n\times\{0\}$ of $\alpha$ 
does not contain the direction $y=(0_n,1)$ of recession of 
$\widetilde{\mathcal{C}}$, i.e.\ a direction $y$ with
$\widetilde{\mathcal{C}}+y\subset \widetilde{\mathcal{C}}$, so 
$K$ is closed, cf.\ Theorem 9.1 in \cite{Rock72}. 
\end{Rem}
\begin{Cor}
\label{cor2:gobyl}
If $\mathcal{P}$ is discrete then the following statements 
are equivalent:
\begin{enumerate}
\item[(i) ] $\mathcal{V}$ is a polyhedron,
\item[(ii) ] $K$ is a polyhedron,
\item[(iii) ] $\mathcal{C}$ is a polytope.
\end{enumerate}
\end{Cor}
\begin{Rem}[Delaunay triangulations]\upshape
\label{rem:geoalg}
A \emph{Delaunay diagram} of a finite generator $\mathcal{P}$ is 
defined as a tessellation of $\conv(\mathcal{P})$ where a circumsphere 
of a cell is an empty sphere. Thereby a circumsphere of a cell is an
\emph{empty sphere} (also \emph{empty circle}) if the interior of the 
corresponding ball has an empty intersection with $\mathcal{P}$. A 
well known construction method for Delaunay diagrams is the
\emph{lifting construction} based on the map
$L : \mathbb{E}^n \rightarrow 
\mathbb{E}^{n+1}, \; x \mapsto (x, \|x\|^2)$. A Delaunay diagram of 
$\mathcal{P}$ is obtained as the orthogonal projection of the 
\emph{lower convex hull} of $L(\mathcal{P})$ onto the 
$(x_1,\dots,x_n)$-plane. That is, the edges of
\[
\{(x,z)\in\conv(\mathcal{P})\times\mathbb{E}\mid
\exists\, y\leq z \text{ such that } (x,y)\in
\conv(L(\mathcal{P}))\}\;\subset\;\mathbb{E}^{n+1}, 
\]
projected orthogonally to the $(x_1,\dots,x_n)$-plane, produce a 
Delaunay diagram for $\mathcal{P}$, see for example Okabe et al.\ 
\cite{OBS00}. In particular, the edges of $\pos(L(\mathcal{P}))$ 
correspond to the edges emanating from $0_n$ in this diagram. 
\par
Heuristically, we consider the lifting construction for an infinite 
discrete generator $\mathcal{P}$. The positive hull of the lifted 
generator is
\[
\pos(L(\mathcal{P}))
\; = \;
\pos\left(\left(\begin{smallmatrix}\inv(x)\\1\end{smallmatrix}
\right),\;x\in\mathcal{P}\setminus\{0_n\}\right)
\; =\;
\bigcup_{\lambda\geq 0}\lambda\left[
\conv(\,\mathcal{R}\,)\times\{1\}\right].
\]
Similarly as in Remark~\ref{rem:compact} we have
$\overline{\pos(L(\mathcal{P}))}=K$, if $\mathcal{P}$ is 
unbounded (otherwise $\mathcal{P}$ is finite). If the 
Voronoi cell $\mathcal{V}$ is not a polyhedron then by 
Theorem~\ref{gobyl} the characteristic cone $K$ is not a polyhedron, 
so $\pos(L(\mathcal{P}))$ is not a polyhedron. This is in 
accordance with the Delaunay diagram having infinitely many edges 
emanating from the origin.
\end{Rem}
\section{The direction cone}
\label{sec:direction_cone}
\par
We compare the convex reciprocal $\mathcal{C}$ to the direction cone 
$\mathcal{D}$. While a polyhedral Voronoi cell $\mathcal{V}$ was found 
equivalent to a polyhedral closure $\overline{\mathcal{C}}$ in the 
last section, we will see in this section that a polyhedral Voronoi 
cell $\mathcal{V}$ is a stronger condition compared to a polyhedral 
direction cone $\mathcal{D}$. In the case of a discrete generator 
$\mathcal{P}$ these conditions are equivalent in accordance with 
Theorem \ref{inas_theorem}.
\par
Let us study what consequences a polyhedral direction cone 
$\mathcal{D}$ can have for the Voronoi cell $\mathcal{V}$. We continue 
in Remark~\ref{rem:gauge} with a proof sketch of a necessary assertion,
omitting to explain the concepts needed for a proof.
\begin{Rem}\upshape
\label{rem:gauge}
If $G\subset\mathbb{E}^n$ is a polyhedron containing the origin $0_n$ 
and if $H=\pos(G)$ then there exists an $\epsilon>0$ such that for the 
open ball 
$B_{\epsilon}(0_n):=\{x\in\mathbb{E}^n\mid\|x\|<\epsilon\}$ we have
\begin{equation}
\label{eq:gauge}
G\,\cap\,B_{\epsilon}(0_n)\;=\;H\,\cap\,B_{\epsilon}(0_n).
\end{equation}
For a proof we can use the \emph{gauge} of $G$ defined for 
$x\in\mathbb{E}^n$ by
\[
\gamma(x)\;:=\;\inf\,\{\,\lambda\geq 0\,\mid\,x\in\lambda C\,\}.
\]
For (\ref{eq:gauge}) to hold it is sufficient to find some 
$\epsilon>0$ such that $\gamma(x)<\epsilon^{-1}$ holds for all $x$ in 
the unit sphere $S(H):=\{x\in H\mid\|x\|=1\}$. The gauge 
$\gamma$ is a positively homogeneous function, whence it has finite 
values on $H$. On the other hand, with $0_n\in G$, the positive hull 
$H=\pos(G)$ is a polyhedron and as such, is locally simplicial. These 
facts can be found in Rockafellar \cite{Rock72}. Theorem 10.2 in this 
reference concludes that $\gamma$ is upper semi-continuous on $H$. 
Since the unit sphere $S(H)$ is compact, $\gamma$ has a finite 
maximum there. This proves (\ref{eq:gauge}).
\par
Asking for sharpness of (\ref{eq:gauge}), let
$G:=\{x\in\mathbb{E}^2\mid\|x-(-\frac{1}{2},0)\|\leq\frac{1}{2}\}$, 
this is  a closed disk touching the origin. Here the gauge $\gamma$ 
still is upper semi-continuous on
$H\setminus\{0_2\}=\{(x,y)\in\mathbb{E}^2|x<0\}$ but the unit sphere 
$S(H)$ is not compact and for each $\epsilon>0$ we have
$G\cap B_{\epsilon}(0_n)\subsetneq H\cap B_{\epsilon}(0_n)$. A related 
example is the convex reciprocal for the example in 
Figure~\ref{fig:bsp3_2_1} with reciprocal
$\mathcal{R}=\{(-1,t)(1+t^2)^{-1}\mid t\in\mathbb{Z}\}$ included in 
the boundary of $G$. 
\end{Rem}
\begin{Pro}
\label{pro:d_to_c}
If the generator $\mathcal{P}$ is discrete and if the direction cone 
$\mathcal{D}$ is a polyhedron, then the convex reciprocal 
$\mathcal{C}$ is a polytope.
\end{Pro}
\par\noindent{\bf Proof.} 
If the cone $\mathcal{D}=\pos(\mathcal{P})$ is finitely generated, then
we can assume that it is finitely generated by points of the generator
$\mathcal{P}$ or by points of the reciprocal $\mathcal{R}$, likewise:
for $r_1,\ldots,r_k\in\mathcal{R}$ we have
\[
\mathcal{D}\;=\;\pos(r_1,\ldots,r_k).
\]
Now we consider the polytope
\[
\widetilde{\mathcal{C}}\;:=\;\conv(0_n,r_1,\ldots,r_k).
\]
Since $\mathcal{D}=\pos(\widetilde{\mathcal{C}})$ we meet the 
assumptions of Remark~\ref{rem:gauge} and can infer that 
$\mathcal{D}\setminus\inv(\widetilde{\mathcal{C}})$ is bounded. So 
this set contains at most finitely many points of the discrete 
generator $\mathcal{P}$. Then all but finitely many points 
$s_1,\ldots,s_l$ of $\mathcal{R}$ belong to $\widetilde{\mathcal{C}}$ 
and therefore the convex reciprocal
$\mathcal{C}=\conv(0_n,r_1,\ldots,r_k,s_1,\ldots,s_l)$ is a polytope.
\hfill $\Box$ \\
\begin{Cor}
\label{cor:disc:d_c:cell}
If the generator $\mathcal{P}$ is discrete and if the direction cone 
$\mathcal{D}$ is a polyhedron then the Voronoi cell $\mathcal{V}$ is a 
polyhedron.
\end{Cor}
\par
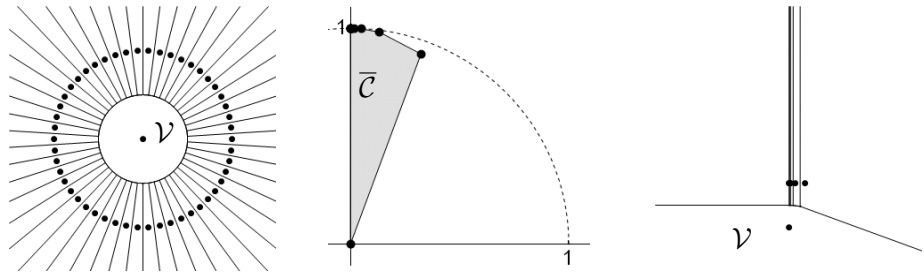
\begin{figure}
\begin{center}
\begin{picture}(12,3.5)
\put(0,0){\includegraphics[height=3.5cm, bb=0 0 500 300]%
{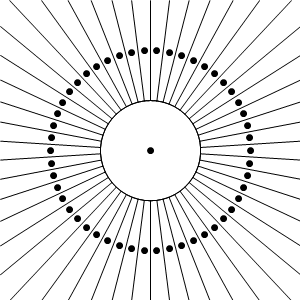}}
\put(4.2,0){\includegraphics[height=3.5cm, bb=0 0 500 300]%
{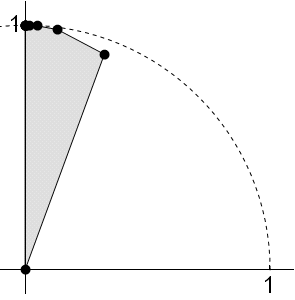}}
\put(8.5,0){\includegraphics[height=3.5cm, bb=0 0 500 300]%
{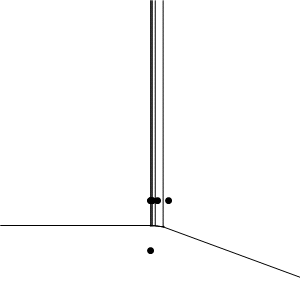}}
\put(1.9,1.7){$\mathcal{V}$}
\put(4.6,2.3){$\overline{\mathcal{C}}$}
\put(9.5,0.3){$\mathcal{V}$}
\end{picture}
\caption{\label{fig:kreis}A finite sketch of the 
Voronoi diagram for unit sphere with center
$\mathcal{P}:=\{(\cos(\varphi),\sin(\varphi))%
\mid\varphi\in[0,2\pi)\}\cup\{0_2\}$ is depicted (left). As the convex 
reciprocal $\overline{\mathcal{C}}=\mathcal{C}=\conv(\mathcal{P})$ is 
not a polyhedron, the Voronoi cell $\mathcal{V}$ is not a polyhedron,
cf.\ Proposition \ref{pro:gobyl}. Another example is the Voronoi cell 
$\mathcal{V}$ 
of $\mathcal{P}:=\{(e^{-x},1)\mid x=1,2,3,\ldots\}\cup\{0_2,(0,1)\}$
(right). Here the convex reciprocal $\overline{\mathcal{C}}$, having 
the infinitely many extreme points $\mathcal{R}$, is not a polyhedron 
(middle). Still, the direction cones are polyhedral in both examples.}
\end{center}
\end{figure}

The above conclusion follows from Proposition \ref{pro:d_to_c} and 
Corollary \ref{cor2:gobyl}, Figure~\ref{fig:par} shows an 
application. Figure~\ref{fig:kreis} demonstrates that a polyhedral 
direction cone is not sufficient for a polyhedral Voronoi cell
without the assumption of the discrete generator. 
\par
Assuming the discrete generator, a polyhedral closure 
$\overline{\mathcal{D}}$ of the direction cone is not sufficient, we 
remember Figure~\ref{fig:bsp3_2_1} and Figure~\ref{fig:hyp}. Now
we will see that a polyhedral closure $\overline{\mathcal{D}}$ of the 
direction cone is necessary regardless of the generator.
\begin{Pro}
\label{pro:c_d}
If the convex reciprocal  $\mathcal{C}$ is a polyhedron then the 
direction cone $\mathcal{D}$ is a polyhedron. If the closure 
$\overline{\mathcal{C}}$ is a polyhedron then the closure 
$\overline{\mathcal{D}}$ is a polyhedron.
\end{Pro}
\par\noindent{\bf Proof.}
We have the trivial chain of inclusions
\[
\mathcal{D}\;=\;\pos(\mathcal{C})\;\subset\;
\pos(\overline{\mathcal{C}})\;\subset\;\overline{\pos(\mathcal{C})}
\;=\;\overline{\mathcal{D}}.
\]
With $\overline{\mathcal{C}}$ being a polyhedron containing the origin,
Corollary 19.7.1 in \cite{Rock72} proves that the positive hull
$\pos(\overline{\mathcal{C}})$ is a polyhedron. Thus, with 
$\pos(\overline{\mathcal{C}})$ being closed we obtain that
$\overline{\mathcal{D}}=\pos(\overline{\mathcal{C}})$ is a polyhedron.
Assuming that $\mathcal{C}$ is a polyhedron, we can replace 
$\mathcal{C}$ by $\overline{\mathcal{C}}$ and obtain 
$\mathcal{D}=\pos(\overline{\mathcal{C}})$.
\hfill $\Box$ \\
\begin{Cor}
\label{cor:c_d:cell}
If the Voronoi cell $\mathcal{V}$ is a polyhedron then the closure 
$\overline{\mathcal{D}}$ of the direction cone is a polyhedron.
\end{Cor}
\begin{Cor}
\label{cor:disc:c_d:cell}
If the generator $\mathcal{P}$ is discrete and if the Voronoi 
cell $\mathcal{V}$ is a polyhedron then the direction cone 
$\mathcal{D}$ is a polyhedron.
\end{Cor}
\par
\begin{figure}
\begin{center}
\begin{picture}(12,3.5)
\put(0,0){\includegraphics[height=3.5cm, bb=0 0 500 300]%
{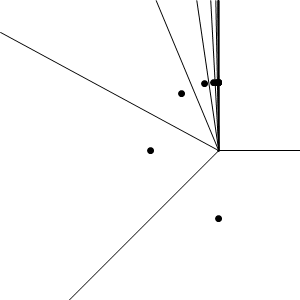}}
\put(4.3,0){\includegraphics[height=3.5cm, bb=0 0 500 300]%
{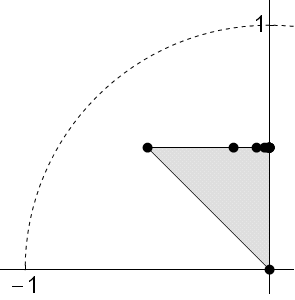}}
\put(8.5,0){\includegraphics[height=3.5cm, bb=0 0 500 300]%
{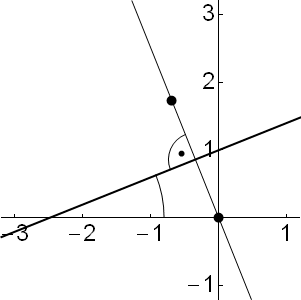}}
\put(2.5,0.1){$\mathcal{V}$}
\put(6.8,1.2){$\overline{\mathcal{C}}$}
\put(9.97,1.07){$\alpha$}
\put(10,2.2){$p_{\alpha}$}
\end{picture}
\caption{\label{fig:klappe}We consider the generator
$\mathcal{P}:=\{0_2,p_{\alpha(1)},p_{\alpha(2)},p_{\alpha(3)},%
\ldots\}$ with
$p_{\alpha}:=2\cos(\alpha)(-\sin(\alpha),\cos(\alpha))$ and 
$\alpha(t):=\frac{\pi}{4}e^{-t}$. The closed convex reciprocal 
$\overline{\mathcal{C}}$ is a triangle (middle), so the Voronoi cell 
$\mathcal{V}$ is a polyhedron by Proposition \ref{pro:gobyl} (left). 
While the direction cone 
$\mathcal{D}=\pos(\{(\lambda,1)\mid -1\leq\lambda<0\})$ is not closed, 
the closure $\overline{\mathcal{D}}$ is a polyhedron by
Corollary \ref{cor:c_d:cell}. The idea to the example 
(right): we narrow $\mathcal{V}$ below certain lines through $(0,1)$.}
\end{center}
\end{figure}
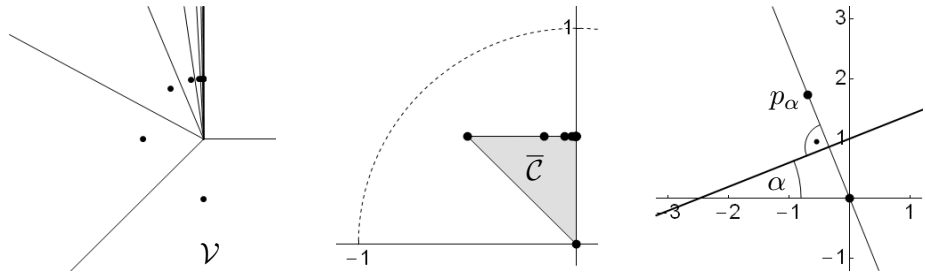

The above conclusions follow from Proposition \ref{pro:c_d} 
together with Proposition~\ref{pro:gobyl} and  
Corollary~\ref{cor2:gobyl} in this order. Figure~\ref{fig:klappe} 
introduces a polyhedral Voronoi cell $\mathcal{V}$ where $\mathcal{D}$ 
is not closed. So, a polyhedral direction cone $\mathcal{D}$ is 
necessary for a polyhedral Voronoi cell $\mathcal{V}$ only in the case 
of a discrete generator.
\section{Bounded cells}
\label{sec:bounded_cells}
\par
Using normal cones we prove a condition when the Voronoi cell 
$\mathcal{V}$ is bounded. For a discrete generator $\mathcal{P}$ this 
is a condition when $\mathcal{V}$ is a polytope.
\par
The \emph{normal cone} of a convex subset $C\subset\mathbb{E}^n$ at 
$x\in C$ is defined by
\[
N(C,x)\;:=\;\{\,u\in\mathbb{E}^n\,\mid
\,\langle y-x,u\rangle\leq 0 \text{ for all }y\in C\,\}.
\]
This is the set of vectors $u\in\mathbb{E}^n$ having no acute angle at 
$x$ with any point $y\in C$. An example is
\begin{equation}
\label{eq:normal}
N(\conv(\mathcal{P}),0_n)\;=\;
N(\mathcal{C},0_n)\;=\;
\{\,u\in\mathbb{E}^n\,|\,\langle y,u\rangle\leq 0
\text{ for all } y\in\mathcal{P}\,\},
\end{equation}
where the set equalities hold because for $u\in\mathbb{E}^n$ the 
inequality $\langle y,u\rangle\leq 0$ for all $y\in\mathcal{P}$ is 
equivalent to this inequality for all $y\in\mathcal{R}$ 
or for all $y$ in the convex hull of one of these sets. The 
inequalities are even equivalent to these with $y$ 
running through the positive hull $\pos(\mathcal{P})=\mathcal{D}$, so 
the \emph{polar} of the direction cone
$\mathcal{D}^*:=\{\,u\in\mathbb{E}^n\,\mid\,\langle y,u\rangle
\leq 0\text{ for all }y\in\mathcal{D}\,\}$ satisfies 
\begin{equation}
\label{eq:polar}
\mathcal{D}^*\;=\;N(\conv(\mathcal{P}),0_n)\;=\;N(\mathcal{C},0_n).
\end{equation}
The \emph{recession cone} of $\mathcal{V}$ describes unbounded 
directions of $\mathcal{V}$, it is
\[
\rec(\mathcal{V})\;:=\;\{\,u\in\mathbb{E}^n\,\mid\, x+\lambda u\in 
\mathcal{V}\text{ for all }x\in\mathcal{V}, \lambda \geq 0\,\}.
\]
\begin{Lem}\label{Lem1}
The equality of cones $N(\conv(\mathcal{P}),0_n)=\rec(\mathcal{V})$ 
holds.
\end{Lem}
\par\noindent
\textbf{Proof.}
We have the representation $2\mathcal{V}\;=\;
\bigcap_{p \in \mathcal{P} \smallsetminus \{ 0_n \}} H^-(p,\|p\|^2)$
by half spaces (\ref{intro:vorocellzero}). With Corollary 8.3.3 in 
\cite{Rock72} the recession cone 
$\rec(\mathcal{V})=\rec(2\mathcal{V})$ becomes
\[
\textstyle
\bigcap_{p\in\mathcal{P}\setminus\{0_n\}}
\rec(H^-(p,\|p\|^2))
\;=\;\bigcap_{p\in\mathcal{P}\setminus\{0_n\}}H^-(p,0).
\]
The proof is completed by (\ref{eq:normal}).
\hfill $\Box$ \\
\par
As a consequence we can determine boundedness of the cell 
$\mathcal{V}$. For $M\subset\mathbb{E}^n$ let $M^{\circ}$ denote the 
interior of $M$ in the topology of $\mathbb{E}^n$. 
\begin{Pro}
\label{pro:bounded}
The Voronoi cell $\mathcal{V}$ is bounded if and only if 
$0_n\in\mathcal{C}^{\circ}$ if and only if
$0_n\in\conv(\mathcal{P})^{\circ}$. If $\mathcal{P}$ is discrete, then
$\mathcal{V}$ is a polytope if and only if one of these equivalent 
conditions holds.
\end{Pro}
\par\noindent
\textbf{Proof.}
By Theorem 8.4 in \cite{Rock72} the Voronoi cell $\mathcal{V}$ is 
bounded if and only if the recession cone $\rec(\mathcal{V})$ is zero. 
Using the equality $N(\conv(\mathcal{P}),0_n)=\rec(\mathcal{V})$ in 
Lemma~\ref{Lem1} this is equivalent to a zero normal cone
$N(\conv(\mathcal{P}),0_n)$. Now by Theorem 13.1 in \cite{Rock72}, the 
normal cone at $0_n$ is zero if and only if
$0_n\in\conv(\mathcal{P})^{\circ}$ holds. Using (\ref{eq:normal}) we 
can argue with $\mathcal{C}$ in place of $\conv(\mathcal{P})$.
In the discrete case, the cell $\mathcal{V}$ is a generalized 
polyhedron, see Remark~\ref{generalized_poly}. But a generalized 
polyhedron is bounded if and only if it is a polytope. 
\hfill $\Box$ \\
\par
The zero normal cone responsible for a bounded Voronoi cell in 
Proposition \ref{pro:bounded} is equivalent to the equality 
$\overline{\mathcal{D}}=\mathbb{E}^n$ through polarity of closed 
cones, cf.\ (\ref{eq:polar}) and Theorem 14.1 in \cite{Rock72}. Since 
a convex cone $\mathcal{D}\subsetneq\mathbb{E}^n$ is included in a 
half space, we have for arbitrary generator $\mathcal{P}$
\[
\mathcal{V}\quad\text{is bounded}
\quad\iff\qquad\mathcal{D}\;=\;\mathbb{E}^n
\quad\iff\qquad\overline{\mathcal{D}}\;=\;\mathbb{E}^n.
\]
\par
\begin{figure}
\begin{center}
\begin{picture}(3.5,3.5)
\put(0,0){\includegraphics[height=3.5cm, bb=0 0 300 300]%
{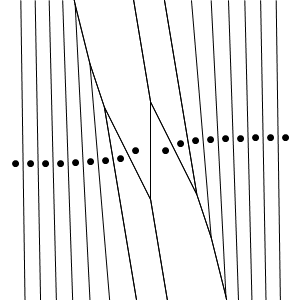}}
\end{picture}
\caption{\label{fig:open_hull}
All Voronoi cells of the generator
$\mathcal{P} := \{ \pm(n, 1-\frac{1}{n}) \mid n \in \mathbb{N} \}$
are polytopes by Proposition~\ref{pro:bounded} as
$\conv(\mathcal{P})=\{(x,y)\in\mathbb{E}^2\mid |y|<1\}$ is open.}
\end{center}
\end{figure}

Figure \ref{fig:open_hull} shows an example of a discrete generator in 
$\mathbb{E}^2$, which is bounded in $y$-direction and where 
nevertheless every Voronoi cell is a polytope. 
%
%
%
%
%
%
%
%
\section{Conclusion}\label{Sec:Con}
By considering a Voronoi cell $\mathcal{V}$ as a problem in linear 
semi-infinite programming, we obtain an equivalent condition when
$\mathcal{V}$ is a polyhedron. This is the condition that the 
closure $\overline{\mathcal{C}}$ of the convex reciprocal is a 
polyhedron and the condition simplifies in the case of a discrete 
generator $\mathcal{P}$ to the condition that the convex reciprocal 
$\mathcal{C}$ is a polytope.
\par
While the polyhedral Voronoi cell $\mathcal{V}$ implies the polyhedral 
closure $\overline{\mathcal{D}}$ of the direction cone, the polyhedral 
cone $\mathcal{D}$ follows only in the discrete case. 
Conversely, the polyhedral direction cone $\mathcal{D}$ implies a 
polyhedral Voronoi cell only in the discrete case.
\par
The closure $\overline{\mathcal{D}}$ gives only advice if the Voronoi 
cell $\mathcal{V}$ is bounded or not.\\
%
%
\par\noindent
\textbf{Acknowledgment.}
Special thanks to Andreas Knauf for general support during our
meeting in Erlangen in August '09, to Oliver Stein who brought 
\cite{Gob98} to our attention and to the anonymous referee, who brought
\cite{Mar04} to our attention.


%
%
%

\begin{thebibliography}{1}
%
%
%
%
\bibitem{Aur02} O.\ Aichholzer and F.\ Aurenhammer, Voronoi diagrams - 
computational geometry's favorite. Special Issue on Foundations of 
Information Processing of TELEMATIK, 1, 7--11 (2002).

%
\bibitem{Gob98} M.\,A.\ Goberna and M.\,A.\ L\'opez,
{\it Linear semi-infinite optimization}, Wiley Series in Mathematical 
Methods in Practice, Chichester (1998).

%
\bibitem{Grub07} P.\,M.\ Gruber, \emph{Convex and Discrete Geometry}, 
Grundlehren der mathematischen Wissenschaften, vol.\ 336, 
Springer, Berlin (2007).

%
\bibitem{Mar04} H.\ Martini and K.\,J.\ Swanepoel,
The geometry of Minkowski spaces -- a survey.\ II., 
Expo.\ Math.\ 22, No.\ 2, 93--144 (2004).

%
\bibitem{OBS00} A.\ Okabe, B.\ Boots, K.\ Sugihara and S.\,N.\ Chiu, 
\emph{Spatial Tessellations, Concepts and Applications of Voronoi 
Diagrams}, Wiley Series in Probability and Statistics, John Wiley \& 
Sons, Chichester (2000).

%
\bibitem{Rock72} R.\,T.\ Rockafellar, \emph{Convex analysis},
Princeton University Press, Princeton (1972).

%
\bibitem{Voigt08} I.\,K.\ Voigt, Voronoizellen diskreter Punktmengen, 
Ph.D.\ thesis, TU Dortmund University, Faculty of Mathematics, 
Dortmund (2008).

%
\bibitem{Zie95} G.\,M.\ Ziegler, {\it Lectures on polytopes},
Graduate Texts in Mathematics 152, Springer, Berlin (1995).
%
\end{thebibliography}
\end{document}